\documentclass[twocolumn]{aastex62}
\received{2018 November 12th}
\revised{2019 January 11th}
\accepted{2019 January 15th}

\usepackage{graphicx}
\usepackage{amssymb}
\usepackage{amsmath}
\usepackage{epstopdf}
\shorttitle{Exocometary CO around M dwarf TWA 7}
\shortauthors{Matr\`a, Luca}

\begin{document}


\title{On the Ubiquity and Stellar Luminosity Dependence of Exocometary CO Gas: Detection around M Dwarf TWA 7}

\author[0000-0003-4705-3188]{L. Matr\`a}
\altaffiliation{Submillimeter Array (SMA) Fellow}
\affil{Harvard-Smithsonian Center for Astrophysics, 60 Garden Street, Cambridge, MA 02138, USA}
\email{luca.matra@cfa.harvard.edu}
\author{K. I. \"{O}berg}
\affil{Harvard-Smithsonian Center for Astrophysics, 60 Garden Street, Cambridge, MA 02138, USA}
\author{D. J. Wilner}
\affil{Harvard-Smithsonian Center for Astrophysics, 60 Garden Street, Cambridge, MA 02138, USA}
\author{J. Olofsson}
\affil{Instituto de F\'isica y Astronom\'ia, Facultad de Ciencias, Universidad de Valpara\'iso, Av. Gran Breta\~na 1111, Valpara\'iso, Chile}
\affil{N\'ucleo Milenio Formaci\'on Planetaria - NPF, Universidad de Valpara\'iso, Av. Gran Breta\~na 1111, Valpara\'iso, Chile}
\author{A. Bayo}
\affil{Instituto de F\'isica y Astronom\'ia, Facultad de Ciencias, Universidad de Valpara\'iso, Av. Gran Breta\~na 1111, Valpara\'iso, Chile}
\affil{N\'ucleo Milenio Formaci\'on Planetaria - NPF, Universidad de Valpara\'iso, Av. Gran Breta\~na 1111, Valpara\'iso, Chile}

\begin{abstract}
Millimeter observations of CO gas in planetesimal belts show a high detection rate around A stars, but few detections for later type stars. We present the first CO detection in a planetesimal belt around an M star, TWA 7. The optically thin CO (J=3-2) emission is co-located with previously identified dust emission from the belt, and the emission velocity structure is consistent with Keplerian rotation around the central star. The detected CO is not well shielded against photodissociation, and must thus be continuously replenished by gas release from exocomets within the belt. 
We analyze in detail the process of exocometary gas release and destruction around young M dwarfs and how this process compares to earlier type stars. Taking these differences into account, we find that CO generation through exocometary gas release naturally explains the increasing CO detection rates with stellar luminosity, mostly because the CO production rate from the collisional cascade is directly proportional to stellar luminosity. More luminous stars will therefore on average host more massive (and hence more easily detectable) exocometary CO disks, leading to the higher detection rates observed.
The current CO detection rates are consistent with a ubiquitous release of exocometary gas in planetesimal belts, independent of spectral type.


\end{abstract}






\section{Introduction}
\label{sect:intro}
Planetesimal belts, also known as debris disks, are extrasolar Kuiper and asteroid belt analogues, detected around a significant fraction of nearby, main-sequence solar-type (FGK) and A stars \citep[for a review, see e.g.][]{Hughes2018, Wyatt2018}. Detectable belts are collision-dominated environments, where the mass of the observed dust and larger planetesimals in a so-called \textit{collisional cascade} depletes over time \citep[as observed, e.g.][]{Wyatt2007b, Holland2017,Sibthorpe2018}. 
To date, 19 of these belts have been detected in one or several gas tracers. Detection of gas in planetesimal belts has the potential to inform us about exocometary compositions \citep{Matra2017b,Matra2018a}, particularly during the $\sim$10-100 Myr period when terrestrial planet formation is still ongoing, and volatile delivery events through exocomets may be commonplace \citep[e.g.][]{Morbidelli2012}.

CO dominates the number of detections (17/19), in part thanks to the unprecedented sensitivity of the Atacama Large Millimeter/submillimeter Array (ALMA). For many systems, it remains debated whether the gas is primordial or secondary in origin, where primordial gas is a long-lived remnant of the protoplanetary disk \citep{Kospal2013}, whereas secondary gas is second-generation like the dust, produced by exocometary gas release \citep[e.g.][]{Zuckerman2012, Matra2015}. 

In five of these belts, the CO masses and line opacities have been well established, and the resulting low CO masses and optically thin CO lines can only be explained by a second generation origin \citep{Dent2014,Marino2016,Marino2017a,Matra2017a,Matra2017b,Booth2018}. The origin of gas for the more massive CO-bearing belts remains unconfirmed, but there are indicators suggesting an absence or a significant depletion of H$_2$, which would point toward a secondary origin for them, too \citep[e.g.][]{Hughes2017, Higuchi2017}. 

A puzzling outcome of the first global analysis of the CO detections is a significantly higher occurrence rate of CO in belts around A stars compared to lower luminosity stars \citep{Moor2017}. Indeed, no detection has been reported to date around stars of type later than F, which constitute the overwhelming majority of stars in our Galaxy. 

In this work, we aim to understand how the presence of CO gas in a planetesimal belt is affected by its stellar host, and explain the origin of the decreasing CO occurrence rates around stars of decreasing stellar luminosity. We do so by presenting the first ALMA detection of CO in a planetesimal belt around an M3-type \citep{Ducourant2014} dwarf, TWA 7. This nearby star \citep[34.0 pc,][]{Gaia2018} is part of the TWA association \citep{Webb1999}, setting its age at $\sim$10 Myr \citep{Bell2015}. Its infrared excess was detected by \textit{Spitzer} \citep{Low2005} and confirmed by \textit{Herschel} \citep{RiviereMarichalar2013}. The first, scattered light image of the outer regions of the belt was obtained by HST \citep{Choquet2016}, and more recent SPHERE data resolve the belt's inner regions showing broad emission from a face-on ring extending from $\sim$20 to at least $\gtrsim70$ au \citep{Olofsson2018}. Unresolved submillimeter emission was detected by the JCMT \citep{Matthews2007}. However, ALMA observations recently showed that the JCMT flux is actually dominated by an unrelated source $\sim6.6\arcsec$ offset from the star (likely a submillimeter galaxy), although the fainter-than-expected belt is also detected and marginally resolved \citep{Bayo2018}. 

We here analyze the CO cube obtained as part of the same ALMA dataset as described in \citet{Bayo2018}. In \S\ref{sect:obs}, we summarize the interferometric observations, including their calibration and imaging. \S\ref{sect:res} presents our analysis of the CO data and a derivation of physical parameters from measured quantities. In \S\ref{sect:disc}, we demonstrate the exocometary origin of the gas, and examine how of the exocometary release process differs around M stars compared to more luminous stars. Finally, we discuss the origin of the stellar luminosity dependence of the CO detection rates in the context of exocometary gas release. We conclude by summarizing our findings in \S\ref{sect:concl}.

\section{Observations}
\label{sect:obs}

TWA 7 was observed by ALMA in Band 7 using the 12m array in 2016. Three observations were obtained, one with 42 antennas in a compact configuration pointed at coordinates RA: 10h42m29.904s Dec: -33d40m17.098s, and two with 36 antennas in a more extended configuration pointed at coordinates RA: 10h42m30.413s Dec: -33d40m16.700s, altogether spanning baselines from 15.1 m to 3.1 km. Further details of the observations can be obtained in \citet{Bayo2018}. Standard calibrations were applied to each dataset using the ALMA pipeline in CASA v5.1.0. A CO visibility dataset was produced covering $\sim$1200, 244.141 kHz-wide channels around the transition frequency (345.796 GHz), with continuum emission subtracted from it using the \textit{uvcontsub} CASA task. For continuum imaging, we considered all 6.9 GHz covered by the 4 spectral windows of each of the observations.

For both CO and continuum, calibrated visibilities from the different observations and configurations were imaged both separately and together using the \textit{tclean} task within CASA, in mosaic mode with multiscale deconvolution. To improve sensitivity to extended emission originating from the belt, we use natural weighting and further apply a 2$\arcsec$ u-v taper to the visibilities prior to imaging. The final CO data cube reaches an RMS sensitivity of 7 mJy/beam in a 244.141 kHz channel (corresponding to 0.21 km/s at 345.796 GHz) for a synthesized beam size of $1.8\arcsec\times 1.7\arcsec$ (62$\times$58 au at the distance of the star from Earth). The continuum image has an RMS noise level of 0.07 mJy/beam for the same synthesized beam size.

\begin{figure}
 \hspace{-7mm}
   \includegraphics*[scale=0.37]{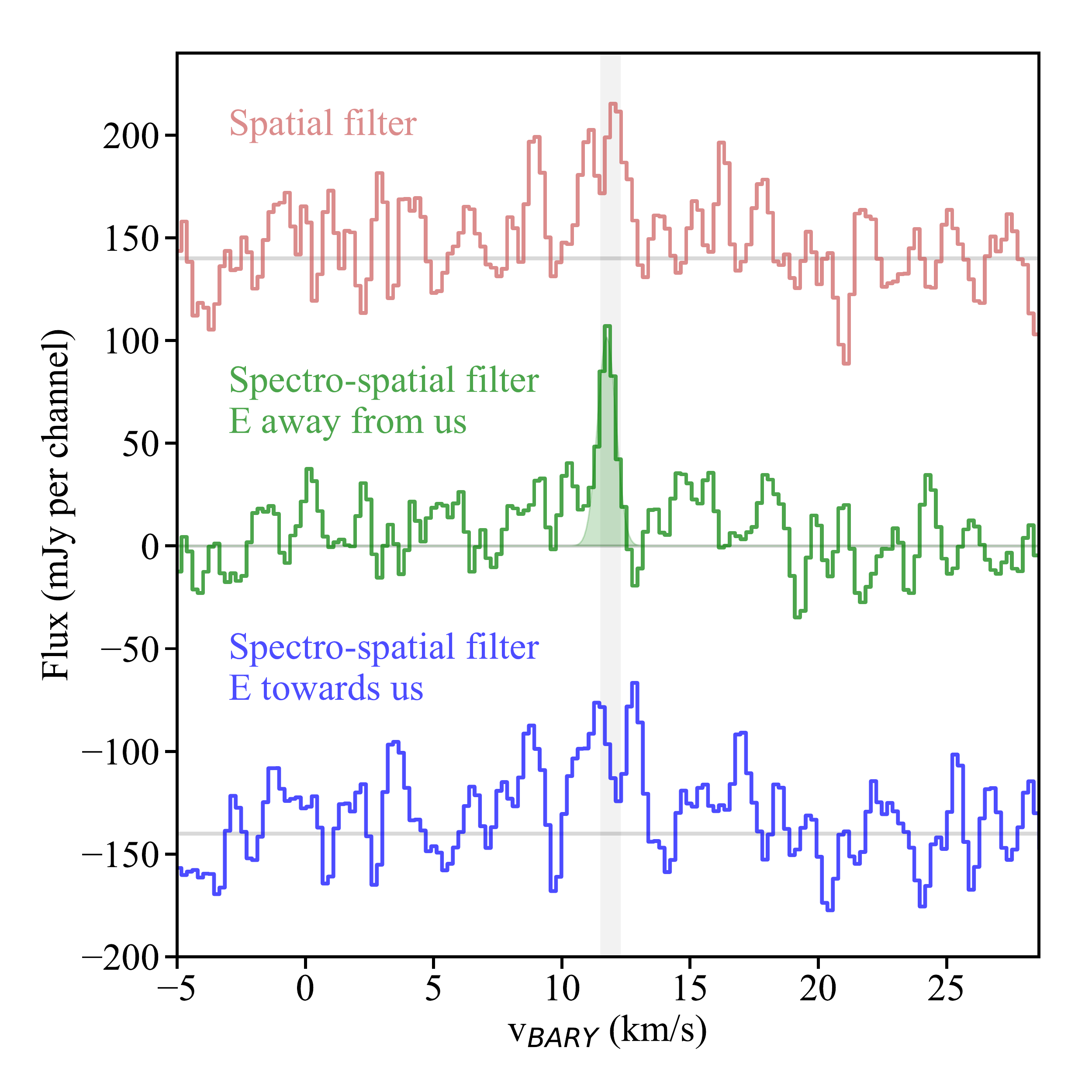}
\vspace{-7mm}
\caption{CO J=3-2 spectra of the belt around TWA 7, spatially integrated over the region where continuum is detected at the $\sim$2$\sigma$ level. The top (red) spectrum is obtained without any spectral filtering, whereas the center and bottom spectra are obtained by shifting the 1D spectra at each pixel location by the negative of their expected Keplerian velocity at that location. Two possible signs of the velocity are possible, depending on whether the East side of the belt is moving away or towards us, leading respectively to the green and blue spectra. The grey region indicates the $\pm1\sigma$ confidence interval for the stellar radial velocity \citep{Gagne2017}. The green shaded region represents the best-fit Gaussian profile to the spectro-spatially filtered emission, used to quantify the integrated line flux.}
\label{fig:1dspectra}
\end{figure}

\section{Results and Analysis}
\label{sect:res}

\subsection{Spectro-spatial filtering: CO J=3-2 detection}
\label{sect:spectrospatial}
No strong emission is observed after quick inspection of the CO data cube around the radial velocity of the star \citep[11.9$\pm$0.4 km/s in the heliocentric frame,][]{Gagne2017}. Following the method of \citet{Matra2015,Matra2017b}, we apply a spectro-spatial filtering technique in order to boost the S/N and achieve maximum sensitivity (Fig. \ref{fig:1dspectra}). First, we simply spatially integrate over the region within $\sim$3.5" from the star where continuum emission from the belt is detected above the 2$\sigma$ level. This produces the red spectrum, where a hint of a Keplerian double-peaked profile already emerges from the noise around the expected stellar velocity. To test whether this signal truly originates from circumstellar gas, we then assume that CO orbits with the Keplerian velocity field expected around a star of 0.55 M$_{\odot}$ \citep{Neuhauser2000}, for a disk with an inclination (13$^{\circ}$) and position angle (PA, 91$^{\circ}$) as derived from high resolution dust imaging in the near-IR \citep{Olofsson2018}. The unknown rotation direction leads to two possible velocity fields, with the East side of the disk moving away from Earth or towards Earth.

For each of the two possible velocity fields, we assign each pixel a given expected radial velocity and shift the 1d spectrum in that pixel along the frequency (velocity) axis, by the negative of its assigned radial velocity. 
Then, we spatially integrate over the region where continuum emission is detected, to produce the green and blue spectra. 
We find that the detection is significantly boosted from a peak S/N of 3.7$\sigma$ to 6.4$\sigma$ for a Keplerian velocity field where the East side of the disk is moving away from us, whereas no boost is achieved when the sign of the velocity field is inverted.

\begin{figure}
 \hspace{-4mm}
   \includegraphics*[scale=0.34]{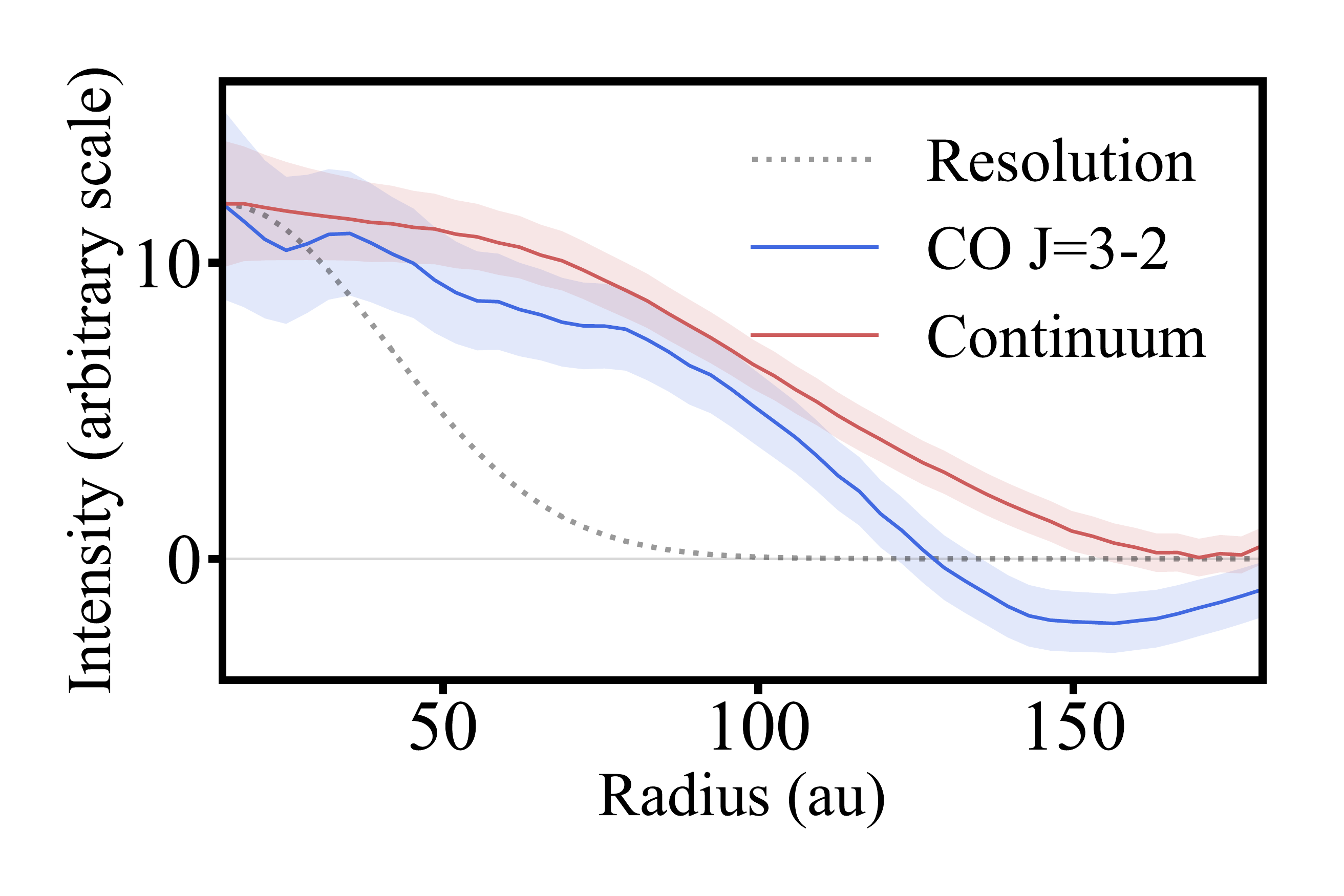}
\vspace{-10mm}
\caption{Radial profiles for continuum emission (red line), and for CO J=3-2 emission after the same spectral filtering that led to the green spectrum (blue line). Shaded regions indicate $\pm1\sigma$ uncertainties, and the dashed line represents the spatial resolution of our observations.} 
\label{fig:radprof}
\end{figure}

After shifting the spectra in each pixel of the cube along the velocity axis, we also produce a moment-0 map by spectrally integrating emission over the 4 channels nearest to the stellar velocity. Then, we extract a radial profile by azimuthally averaging emission in concentric elliptical annuli (accounting for the belt's inclination and PA). This radial profile is shown by the blue shaded region in Fig. \ref{fig:radprof}. We find that CO emission is detected out to $\sim$100 au from the star, with a radial profile consistent with that extracted from the continuum images, tracing the dust emission (red shaded region). The present dataset does not allow us to resolve the inner hole of the dust belt, found to lie at $\sim$20 au in near-IR observations \citep{Olofsson2018}, while at the same time retaining sensitivity to the extended belt structure. 
Future, higher sensitivity observations are necessary for a detailed comparison of the radial structure of both large, planetesimal-tracing grains and CO emission.

We conclude that 1) CO J=3-2 emission is significantly detected at the radial velocity of TWA 7, at a location consistent with that of the belt of mm grains, and 2) its radial velocity field is consistent with circumstellar gas co-located with the dust and in Keplerian rotation around the star, with the East side of the gas disk moving away from Earth.

\subsection{CO mass and optical depth estimate}
\label{sect:massandtau}

We measure the integrated CO line flux by fitting a Gaussian to the green, spectrospatially filtered spectrum of Fig. \ref{fig:1dspectra}, obtaining a value of 91$\pm$20 mJy km/s for a centroid radial velocity of 11.75$\pm$0.08 km/s. Uncertainties were calculated from \textsc{SciPy}'s \textit{curve\_fit} non-linear least squares method \citep{Jones2001}. These rely on the input uncertainty on the flux for each channel, which was assumed to be equal to the RMS of the spectrum multiplied by a factor $\sqrt{2.667}$. The latter was introduced to account for the fact that neighbouring channels are correlated, and the effective bandwidth of the flux in each channel is 2.667 times the width of that channel\footnote{\url{https://safe.nrao.edu/wiki/pub/Main/ALMAWindowFunctions/Note_on_Spectral_Response.pdf}}. Multiplying the RMS by a factor $\sqrt{2.667}$ then takes care of the fact that while the line is resolved over $\sim$7 channels, there are really only $\sim$7/2.667=2.62 independent measurements, therefore ensuring that the error on the final parameters is not underestimated. The uncertainty on the line flux included an extra 10\% added in quadrature to account for ALMA's flux calibration uncertainty in Band 7\footnote{\url{https://almascience.nrao.edu/documents-and-tools/cycle6/alma-technical-handbook}}.

Assuming optically thin emission, we use the non-LTE excitation code of \citet{Matra2015,Matra2018a} to find that the best-fit line flux corresponds to CO masses in the range 0.8-80$\times10^{-6}$ M$_{\oplus}$, where the range comes from the two limiting regimes of molecular excitation, at high collider density (collision-dominated regime, or LTE) and low collider density (radiation-dominated regime). Note that in order to calculate the effect of UV and IR pumping on CO excitation, we adopt the interstellar radiation field (ISRF) of \citet{Draine1978} and a PHOENIX stellar photospheric model fitted to observations of TWA 7 with synthetic photometry (T$_{\rm eff}$=3394 K, log(g)=3.7, [M/H]=0.0), and scaled to a distance of 60 au from the star. If instead we were to scale the model to the inner edge of the belt, or if the model took into account excess UV chromospheric emission (see \S\ref{sect:cophoto}), UV pumping would be stronger and act to reduce the upper limit of the range of masses given above.

We then verify the optically thin assumption by assuming CO emission is co-located with continuum emission (which is informed by the radial profiles in Fig. \ref{fig:radprof}) and that this is co-located with dust emission seen at higher resolution at optical wavelengths \citep{Olofsson2018}. Specifically, we assume the disk to extend between 20 and 100 au, and to have a constant surface brightness. For simplicity, we assume the belt to be perfectly face-on, where this introduces only a $\sim$3\% uncertainty in the length of the column along the line of sight to Earth, given the known belt inclination of $\sim$13$^{\circ}$ \citep{Olofsson2018}. 
Then, the CO mass derived above corresponds to an average column density between $\sim$0.15-15$\times10^{14}$ molecules cm$^{-2}$, which can be used to derive a maximum optical depth of $\tau<0.27$ \citep[using Eq. 3 in][, and the full range of excitation conditions used to derive CO masses]{Matra2017a}. This confirms the validity of our optically thin emission assumption and of our CO mass estimate of 0.8-80$\times10^{-6}$ M$_{\oplus}$.

\section{Discussion}
\label{sect:disc}

\subsection{CO photodissociation around young M dwarfs due to chromospheric FUV emission}
\label{sect:cophoto}


The CO gas detected is continuously affected by the impinging stellar and interstellar UV radiation field, which in the $\sim$900-1100 \AA\ wavelength range leads to photodissociation \citep[e.g.][]{Visser2009}.
Contrary to the conclusion one would draw from a simple comparison of stellar photospheric models, the photodissociation-driving far-UV (FUV) intensity of young M dwarfs is not negligible compared to, for example, coeval A-stars. This is because FUV emission is dominated by emission lines originating from the chromosphere and/or transition region of the star, with only a negligible contribution from the stellar photosphere \citep{Linsky2017}. These lines are strongest around active stars, such as young M dwarfs, but can also be present and dominate the FUV continuum around more massive stars such as $\beta$ Pictoris \citep{Deleuil2001, Bouret2002}. Then, photodissociation of exocometary CO across stars of different luminosities does not simply scale with the stellar effective temperature, but will strongly depend on the details of the stars' chromospheric emission and FUV spectra.
Since TWA 7 has not been observed in the $\sim$900-1100 \AA\ range, we use the spectrum of the similar, young M dwarf AU Mic and rescale its intensities by the ratio of FUV luminosities between 1250 and 1700 \AA, observed for both stars with HST. In the latter UV range, TWA 7 was found to be more luminous than AU Mic by a factor of $\sim$1.5 \citep{Yang2012}.

Since CO photodissociation is a line process, it is sensitive to the presence of stellar chromospheric lines that overlap with electronic CO transitions leading to pre-dissociative excited states.
In the case of $\beta$ Pic, the broad \ion{C}{3} chromospheric line at 977.02 \AA\ is by far the strongest component of the FUV spectrum, and it overlaps with a pre-dissociative absorption band of CO at 977.40 \AA\ \citep[Band 26 in Table 1 of][]{Visser2009}. It therefore dominates the stellar CO photodissociation \citep[see][Fig. 3]{Matra2018a}. The same line carries most of the FUV flux of the coeval ($\sim$23 Myr-old) M-dwarf AU Mic \citep{Redfield2002}.
 The chromospheric line and the CO line do not overlap perfectly, however, and because chromospheric lines are relatively narrow in M dwarfs, only some of the emission contributes to CO photodissociation. Even so, this line is the likely dominant contributor to stellar CO photodissociation around AU Mic, and by inference around TWA 7. We therefore use the \ion{C}{3} stellar line intensities as measured in FUSE spectra at 977.40 \AA\ as a proxy for the strength of CO-photodissociating stellar radiation. For AU Mic, we use the quiescent flux of $\sim0.5\times10^{-13}$ erg cm$^{-2}$ s$^{-1}$ \AA$^{-1}$, but note that this increased to $\sim8.5\times10^{-13}$ erg cm$^{-2}$ s$^{-1}$ \AA$^{-1}$ during an observed flare \citep[see Fig. 17 of][]{Redfield2002}.

Combining AU Mic's 977.40 \AA\ flux with $\beta$ Pic's measurement of $\sim2\times10^{-13}$ erg cm$^{-2}$ s$^{-1}$ \AA$^{-1}$ \citep{Bouret2002}, and scaling to TWA 7 accounting for the respective distances from Earth, we find that the CO-photodissociating radiation of M star TWA 7 is at least $\sim$6\% that of A star $\beta$ Pic. Given that $\beta$ Pic's unshielded CO photodissociation timescale at 85 au is 70 years \citep{Cataldi2018}, TWA 7's star-driven CO photodissociation timescale at 60 au will be 580 years, which is longer than the photodissociation timescale driven by the interstellar radiation field alone \citep[120 years, e.g.][]{Heays2017}. Combining the (quiescent) stellar and interstellar radiation fields, we conclude that the CO unshielded photodissociation timescale around TWA 7 is 99 years at 60 au, varying from 55 years at 25 au to 112 years at 100 au. If flares were taken into account, this timescale could be even shorter.


\subsection{Exocometary origin for TWA 7 CO}
\label{sect:exocomorig}
The low, optically thin levels of CO detected around the $\sim$10 Myr-old star TWA 7 supports a picture where the gas is of secondary origin, i.e. produced within a destructive collisional cascade. But can we rule out that the CO is primordial, or in other words left over from a depleted, old protoplanetary disk? 
Unshielded CO will be photodissociated in $\sim$100 years (\S\ref{sect:cophoto}), but can 
self-shielding or shielding by other gas species have prolonged its lifetime and allow it to survive since the protoplanetary phase of evolution?

A CO molecule lying in the midplane of the disk will see a CO column density of at most $\sim7.5\times10^{14}$ cm$^{-2}$ above the midplane (half of the maximum value derived in \S\ref{sect:massandtau}). Using the self-shielding factors of \citet{Visser2009}, CO can only prolong its own lifetime by a factor 2.5, corresponding to a lifetime of 250 years (see their Table 7). If the disk was primordial in origin, we may expect large amounts of H$_2$ to be present. Conservatively assuming CO to be depleted with respect to H$_2$ compared to interstellar abundances, with a CO/H$_2$ ratio of 10$^{-6}$ \citep[as in the coeval TW Hya protoplanetary disk,][]{Favre2013}, the photodissociation lifetime is only prolonged by another factor 4 to 1000 years, and less so if the CO/H$_2$ ratio was interstellar. Since this is much shorter than the age of the system, we conclude that primordial CO, originating in the protoplanetary disk, is not responsible for the observed CO in the TWA 7 belt (unless the protoplanetary disk dispersed within the last 1000 years, which is very unlikely). The observed CO gas most likely originates in the collisional cascade that also produces the observed dust, i.e. it has an exocometary origin.

\subsection{The exocometary gas release process \\around M dwarfs}
\label{sect:compupdate}
TWA 7 joins a group of 5 other planetesimal belt systems with CO gas confirmed to be of exocometary origin, comprising $\beta$ Pictoris \citep{Matra2017a}, HD181327 \citep{Marino2016}, $\eta$ Corvi \citep{Marino2017a}, Fomalhaut \citep{Matra2017b} and HD95086 \citep{Booth2018}. 
In these disks, the measured CO gas mass can be used to probe the volatile composition of exocomets. This is done by considering a model where the gas is continuously released as part of the collisional cascade, which also produces the observed dust \citep[e.g.][]{Zuckerman2012,Matra2015}. The CO release is assumed to be at steady state, with the release rate being equal to the photodissociation rate. The CO release rate is simply the rate at which solid mass is lost from the collisional cascade, combined with the CO (+CO$_2$) exocometary mass fraction. This allows the exocometary mass fraction to be derived from simple dust, CO and host star observables \citep[Eq. 2 of ][]{Matra2017b}.

\subsubsection{Mass loss rate of small grains}
\label{sect:masslosswind}

In the argument above, extracting the exocometary mass fraction requires knowledge of the mass loss rate of the smallest grains, which for typically observed belts occurs by radiation pressure from the central star. 
However, TWA 7 is an M dwarf, which means that radiation pressure is unable to eject and remove the smallest grains from the system, and stellar winds play a dominant role \citep[e.g.][]{Plavchan2005, AugereauBeust2006}. Stellar winds create outward ram pressure on the grains, an effect analogous to radiation pressure, as well as corpuscular drag causing the grains to lose angular momentum and drift inward, analogously to P-R drag. \citet{StrubbeChiang2006} considered these effects in detail, and showed that belts where the mass loss is dominated by corpuscular wind drag should present an inner hole that is filled with small grains spiralling inward, and a steep outer slope of the surface density distribution. These are not observed in SPHERE images of TWA 7, tracing the smallest grains \citep{Olofsson2018}. We therefore postulate that small grain removal from the TWA 7 belt is dominated by outward stellar wind pressure. 



In a steady-state cascade, the mass loss rate of the smallest, blow-out grains is the same as the total mass of these smallest grains multiplied by the collision rate of grains just above this size, which can be calculated \citep[see][for details]{Matra2017b}. The minimum grain size influences the total mass and area available for collisions, thus impacting the calculated mass loss rate of blow-out grains.
Therefore, the difference between stellar winds for M dwarfs and radiation pressure for more massive stars enters the calculation of the mass loss rate only by setting the size of the smallest grains $D_{\rm min}$. For stellar wind dominated removal, this reads \citep{AugereauBeust2006}
\begin{equation}
D_{\rm min}=2.8\times10^{11}\frac{\dot{M}_{\star}v_{\rm sw}C_{\rm D}}{M_{\star}\rho},
\end{equation}
where $D_{\rm min}$ is in $\mu$m, $\dot{M}_{\star}$ is the mass loss rate of the star due to the wind (in M$_{\odot}$yr$^{-1}$), $v_{\rm sw}$ is the wind velocity (in km/s), C$_{\rm D}$ is the drag coefficient of the free molecular flow, M$_{\star}$ is the stellar mass (in M$_{\odot}$), and $\rho$ is the density of the grains (in kg m$^{-3}$). 

Unfortunately, there is no measurement of these M-dwarf wind properties. However, given that our calculation of the mass loss rate of grains only relies on the minimum grain size the wind produces, we can simply use the $D_{\rm min}$ value of $\sim$0.16 $\mu$m constrained from SED fitting \citep[for astrosilicate grains with $\rho=3500$ kg m$^{-3}$, and an assumed \citet{Dohnanyi1969} size distribution,][]{Bayo2018}. Assuming these grains dominate the observed IR luminosity, and taking TWA 7's stellar mass of 0.51 M$_{\odot}$, fractional luminosity of 1.7$\times$10$^{-3}$, and approximate belt radius of $\sim$60 au with width of $\sim$80 au, we obtain a mass of the smallest grains of $1\times10^{-4}$ M$_{\oplus}$ and a collisional timescale of $3\times10^4$ yr \citep[Eq. 16 and 20 in Appendix B,][]{Matra2017b}. Together, this implies a mass loss rate of small grains from the collisional cascade of 0.003 M$_{\oplus}$ Myr$^{-1}$.


\subsubsection{Estimate of exocometary compositions in TWA 7}

Given a measured CO mass in the range 0.8-80$\times10^{-6}$ M$_{\oplus}$, and taking the unshielded photodissociation timescale of 100 years at 60 au, we estimate a mass loss rate of CO gas through photodissociation of 0.008-0.8 M$_{\oplus}$ Myr$^{-1}$. In steady state, this will correspond to the mass loss rate $\dot{M}_{\rm CO(+CO_2)}$ of CO(+CO$_2$) from the ice.
Adding this to the mass loss rate $\dot{M}_{D_{\rm min}}$ of grains from the bottom of the collisional cascade of 0.003 M$_{\oplus}$ Myr$^{-1}$ (\S\ref{sect:masslosswind}), and assuming that the release rate of volatiles other than CO and its potential parent species CO$_2$ \citep[which are yet to be detected in gaseous form within exocometary belts, e.g.][]{Matra2018a} is negligible, we obtain the total rate of mass loss through the collisional cascade $\dot{M}=\dot{M}_{\rm CO(+CO_2)}+\dot{M}_{D_{\rm min}}$. As long as all CO(+CO$_2$) ice is lost \citep[through e.g. sublimation and/or photodesorption, see discussion in][]{Matra2017b} by the time a large body is ground down to blow-out size grains, then $\dot{M}_{\rm CO(+CO_2)}=f_{\mathrm{CO}+\mathrm{CO}_2}\dot{M}$, where $f_{\mathrm{CO}+\mathrm{CO}_2}$ is the CO(+CO$_2$) mass fraction in TWA 7's exocomets. This reasoning leads to Eq. 1 in \citet{Matra2017b} and allow us to derive a CO(+CO$_2$) mass fraction $f_{\mathrm{CO}+\mathrm{CO}_2}$ $\geq70$\% for exocomets around TWA 7.

This value is high compared to the few to few tens of percent observed for other stars hosting exocometary CO (\S\ref{sect:compupdate}) as well as Solar System comets \citep{Matra2017b}, and would imply icy bodies almost entirely composed of CO and CO$_2$ ice. It is possible that exocomets around M dwarfs do have a distinct composition. However, our estimate is subject to uncertain assumptions that need to be observationally tested. First, the observational estimate of the minimum grain size from the SED could vary by as much as an order of magnitude by changing the assumed grain composition. Second, the photodissociation timescale could vary by a factor of a few from our basic rescaling using the 977.40 \AA\ line flux of AU Mic. This could be due to, for example, observations when the stars were at different activity levels, or if our rescaling doesn't apply to the chromospheric 977.40 \AA\ \ion{C}{3} line. The presence of flares (both in AU Mic and and TWA 7) could also shorten, whereas self-shielding could lengthen, the photodissociation timescale, making the derived CO mass fraction respectively larger or smaller.

Finally, shielding by other atoms or molecules could be more important than assumed in our calculations. Although reasonable H$_2$ abundances cannot reasonably provide sufficient shielding against photodissociating UV light (\S\ref{sect:exocomorig}), atomic carbon produced by CO photodissociation could \citep[e.g.][]{Matra2017a}. Depending on the true CO(+CO$_2$) ice mass fraction in TWA 7's exocomets and on the $\alpha$ viscosity of the gas disk, TWA 7 could be producing C at a fast enough rate for it to shield CO before spreading radially \citep[][e.g. Fig. 18, bottom right]{Kral2018}. This would prolong the CO survival timescale against photodissociation, potentially reducing the derived CO(+CO$_2$) mass fraction to more commonly observed values. Follow-up ALMA measurements of the \ion{C}{1} gas mass and of another CO transition to better constrain the CO gas mass will have the ability to set more stringent constraints on the composition of TWA 7's exocomets.



\subsection{On the stellar luminosity dependence and ubiquity of exocometary CO gas release}
\label{sect:ubiquity}

There are 17 CO detections in extrasolar planetesimal belts, 6 of which (including TWA 7) are conclusively of exocometary origin. These CO-producing exocometary belts span ages between $\sim$10 Myr and 1-2 Gyr, implying that the presence of gas is not limited to young stars. It is indeed possible that \textit{all icy planetesimal belts host exocometary gas at some level}, and that the CO outgassing is simply proportional to the mass loss rate of the collisional cascade. 

\subsubsection{The mass of exocometary gas released depends on the host star's luminosity}
This statement has to be reconciled with the dependence of the detection rate of CO in planetesimal belts on stellar luminosity. \citet{Moor2017} reported a very high detection rate of 68.8$^{+8.9}_{-13.1}$\% in belts around A stars with high fractional luminosity ($>5\times10^{-4}$). However, they also note that the rate drops significantly to $6.7^{+12.5}_{-2.2}$\% for belts around the FG stars in their sample. Can exocometary release explain this trend by producing higher CO masses around more luminous stars?

Inverting Eq. 2 in \citet{Matra2017b}, the CO mass (in M$_{\oplus}$) predicted for any given planetesimal belt reads:
\begin{equation}
\begin{split}
M_{\rm CO}=\dot{M}_{D_{\mathrm{min, rp}}}t_{\mathrm{phd}}\frac{f_{\mathrm{CO}+\mathrm{CO}_2}}{1-f_{\mathrm{CO}+\mathrm{CO}_2}} \\
=1.2\times10^{-3}R^{1.5}\Delta R^{-1}f^2L_{\star}M_{\star}^{-0.5}t_{\mathrm{phd}}\frac{f_{\mathrm{CO}+\mathrm{CO}_2}}{1-f_{\mathrm{CO}+\mathrm{CO}_2}}
\end{split}
\end{equation}
where $R$ and $\Delta R$ are the radius and width of the belt in au, $L_{\star}$ and $M_{\star}$ are the stellar luminosity and mass in L$_{\odot}$ and M$_{\odot}$, $t_{\mathrm{phd}}$ is in years, $f$ is the fractional luminosity of the belt, $\dot{M}_{D_{\mathrm{min, rp}}}$ is the mass loss rate of blow-out grains via radiation pressure, and the resulting CO mass $M_{\rm CO}$ is in M$_{\oplus}$.

This is for the case where the minimum grain size of the cascade is set by radiation pressure, which will apply for stars of spectral type earlier than M. For M stars, the minimum size is set by stellar winds, which causes the expression above to turn into
\begin{equation}
\begin{split}
M_{\rm CO}=\dot{M}_{D_{\mathrm{min, sw}}}t_{\mathrm{phd}}\frac{f_{\mathrm{CO}+\mathrm{CO}_2}}{1-f_{\mathrm{CO}+\mathrm{CO}_2}} \\
=1.6\times10^{5}R^{1.5}\Delta R^{-1}f^2\dot{M}_{\star}v_{\mathrm{sw}}C_{\mathrm{D}}M_{\star}^{-0.5}t_{\mathrm{phd}}\frac{f_{\mathrm{CO}+\mathrm{CO}_2}}{1-f_{\mathrm{CO}+\mathrm{CO}_2}},
\end{split}
\end{equation}
where $\dot{M}_{D_{\mathrm{min, sw}}}$ is the mass loss rate of blow-out grains via stellar winds, $\dot{M}_{\star}$ is the mass loss rate of the stellar wind itself (in M$_{\odot}$yr$^{-1}$), $v_{\rm sw}$ is the wind velocity (in km/s), and C$_{\rm D}$ is the drag coefficient of the free molecular flow. The main difference between the radiation pressure case (Eq. 2) and the stellar wind case (Eq. 3) is that the stellar luminosity is substituted by the product of the stellar wind mass loss rate, its velocity, and the drag coefficient $C_{\rm D}$.

We immediately observe an explicit linear $L_{\star}$ dependence in Eq. 2. This arises from the fact that for the same fractional luminosity (or total dust cross-sectional area), the blow-out size of grains in the cascade is larger and hence the belt's dust mass $M_{\rm dust}$ in the smallest grains is higher around more luminous stars \citep[e.g. Eq. 16 in][]{Matra2017b}. The grain collision rate $R_{\rm col}$ is dependent on the grains' cross-sectional area and hence on the belt's fractional luminosity, which we are keeping constant. Then, the mass loss rate ($=M_{\rm dust}R_{\rm col}$) and therefore the CO mass in Eq. 2 carries a luminosity dependence, because more luminous stars will have larger grains and hence, for a constant fractional luminosity, will host more massive belts.

\begin{figure*}
 \hspace{-7mm}
   \includegraphics*[scale=0.46]{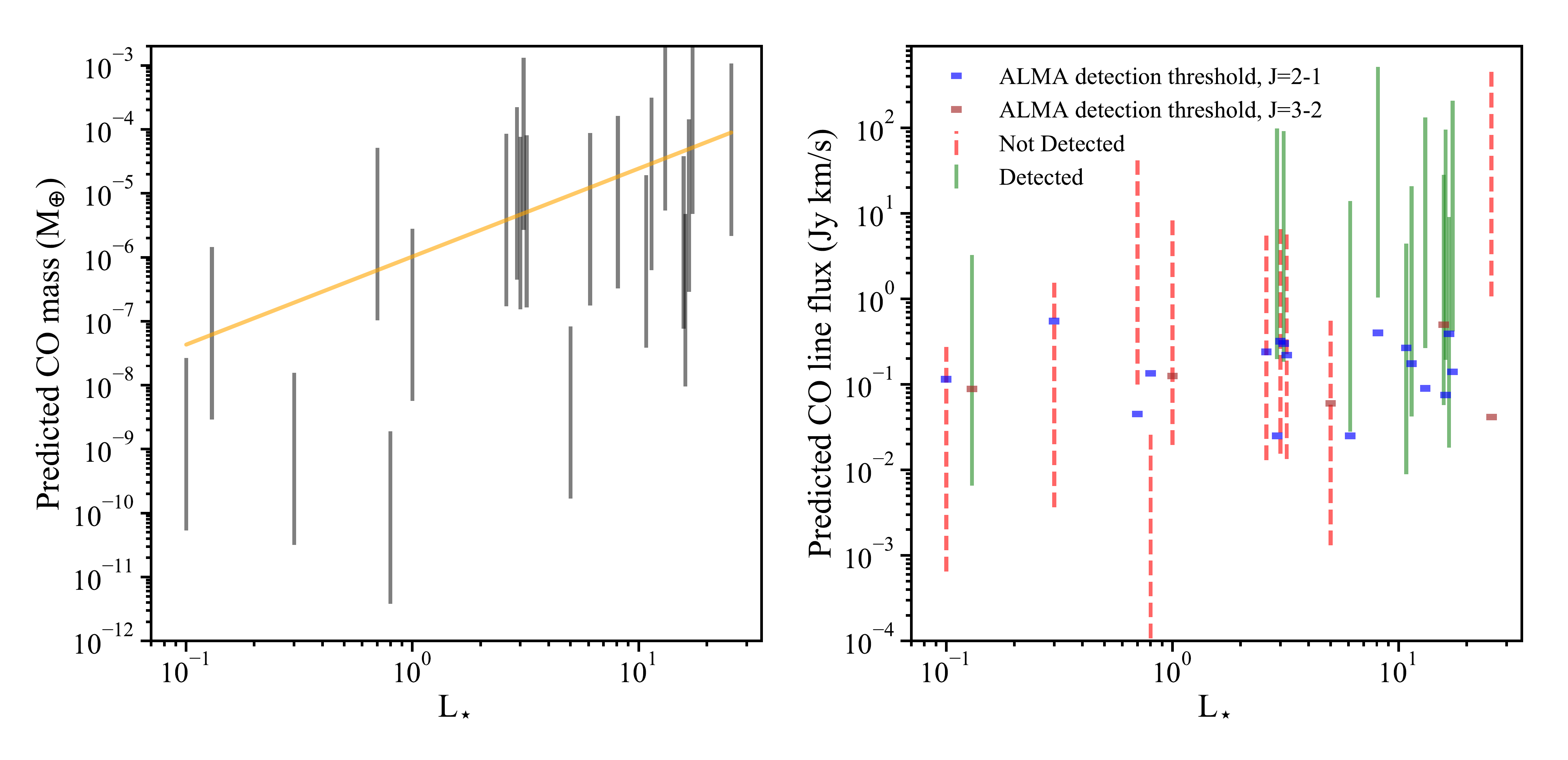}
\vspace{-11mm}
\caption{\textit{Left:} Predicted exocometary CO gas mass for belts which have been resolved and searched for CO by ALMA so far, as a function of stellar luminosity, and for a 0.8-80\% range of assumed CO(+CO$_2$) exocometary ice mass fractions (grey bars). The sample of stellar and belt properties used in the calculation are from this work for TWA 7, from \citet{Kennedy2018} for HR4796A and from \citet{Matra2018b} for the remainder of the sample. The orange line is a prediction assuming $R=73L_{\star}^{0.19}$ \citep{Matra2018b}, a constant photodissociation timescale of 120 years, a CO(+CO$_2$) mass fraction of 20\%, a constant $\Delta R/R=0.4$, $M_{\star}=L_{\star}^{0.25}$, and $f=(1.56\times10^{-3})L_{\star}^{0.2}$. \textit{Right:} Predicted integrated CO line fluxes for the given CO mass ranges in the left panel, given the star's respective distances from Earth and an excitation temperature of 20 K. Green solid bars represent targets with detected CO, whereas red dashed bars are for targets where CO was not detected. Horizontal bars represent, for each target, the 5$\sigma$ detection threshold of its actual ALMA observation from the literature. Blue bars are for targets where the J=2-1 transition was observed, and brown ones are for the J=3-2 transition. In both panels, the exocometary release model predicts increasing masses and fluxes of CO for stars of increasing luminosity. Combined with the detection thresholds, this explains the increasing CO detection rate as a function of stellar luminosity reported by \citet{Moor2017}.} 
\label{fig:predoccurrencerate}
\end{figure*}

Ultimately, the relation between the exocometary CO gas mass $M_{\rm CO}$ and $L_{\star}$ will also incorporate the luminosity dependence of other variables in Eq. 2. To estimate the latter, we consider the sample of belts that have been resolved by ALMA \citep{Matra2018a} and searched for CO. Assuming $R=73L_{\star}^{0.19}$ \citep{Matra2018b}, a constant photodissociation timescale of 120 years, a CO(+CO$_2$) mass fraction of 20\%, a constant $\Delta R/R=0.4$ (the average fractional width of the sample), $M_{\star}=L_{\star}^{0.25}$, and $f=(1.56\times10^{-3})L_{\star}^{0.2}$ (the best-fit from a simple regression of the observed $f-L_{\star}$ relation for the above sample), we find that we should expect $M_{\rm CO}\propto L_{\star}^{1.37}$ (orange line in Fig. \ref{fig:predoccurrencerate}, left panel).

At the same time, we plot the predicted exocometary CO gas mass when taking into account the true $[f,R, \Delta R]$ of each of the belts in the sample (grey bars in Fig. \ref{fig:predoccurrencerate}, left), which will more accurately reflect the scatter in the parameters going into Eq. 2 and 3. In this case, the CO mass is calculated through Eq. 2 for stars with $L_{\star}\geq0.2$ L$_{\odot}$. For TWA 7, we use the mass loss rate from the collisional cascade $\dot{M}_{D_{\mathrm{min, sw}}}$ derived in \S\ref{sect:masslosswind}, while for AU Mic we rescale TWA 7's mass loss rate according to the different stellar and belt parameters that enter Eq. 3.

For each belt, we show the predicted range of exocometary CO gas masses for CO(+CO$_2$) ice mass fractions of 0.8-80\%; we also assume a photodissociation timescale of 120 years (as expected if the interstellar UV radiation field were to dominate over the star's) for all belts except HR4796A, which is a belt with a small radius $R$ around the most luminous star in our sample, therefore clearly subject to much faster photodissociation \citep[$t_{\rm phd}$=8 years,][]{Kennedy2018}. As expected, Fig. \ref{fig:predoccurrencerate} (left) shows that more luminous stars will, on average, host belts with a higher exocometary CO mass.

\subsubsection{From mass to observed line fluxes: comparing with existing data}
Having established that exocometary gas release should produce an increased exocometary CO mass around more luminous stars, we then assess whether this can produce the observed trend of higher detection rates as a function of stellar luminosity. To do so, we use CO masses (Fig. \ref{fig:predoccurrencerate}, left) to compute CO line fluxes (Fig. \ref{fig:predoccurrencerate}, right) for the given line (J=2-1 or J=3-2) that was observed for each belt.

For simplicity we assume, for all belts, optically thin emission for an excitation temperature of 20 K \citep[][Eq. 2]{Matra2015}, which is in between the excitation temperatures (of 9-32 K) measured in belts where a CO line ratio was available \citep{Kospal2013,Matra2017a,Hughes2017}. 
Then, these predicted CO fluxes can be compared to the 5$\sigma$ detection threshold actually achieved by ALMA observations of each target (horizontal bars; blue for J=2-1, pink for J=3-2). These were obtained from the literature \citep{Daley2019subm, Booth2017, Marino2017a,Marino2018a,Lieman-Sifry2016,Marino2016,Booth2018,Matra2017a,Kospal2013,Moor2017, Matra2017b,Kennedy2018}\footnote{For reported detections, we took the uncertainty on the integrated line flux, and subtracted the 10\% ALMA flux calibration uncertainty in quadrature. For HD61005, the upper limit is not reported in \citet{Olofsson2016}, so we calculated the uncertainty on the integrated line flux from the dataset directly as done for TWA 7 here. For the sources in \citet{Lieman-Sifry2016}, the spatially and spectrally integrated uncertainty is measured assuming two different spatial apertures; we rescale all uncertainties to a common 2$\arcsec$ aperture.}. 

Fig. \ref{fig:predoccurrencerate} (right) shows that, on average, the predicted CO flux increases as a function of stellar luminosity. This is as expected from the CO mass trend in the left panel, although with a shallower slope largely caused by the fact that, on average, higher luminosity stars lie further away from Earth. In contrast with the predicted CO fluxes increasing with stellar luminosity, the ALMA detection thresholds remain roughly constant throughout the luminosity range, reflecting the fact that all observations reached 5$\sigma$ line flux sensitivities within less than an order of magnitude of 0.1 Jy km/s.

The predicted CO flux \textit{increasing} and the ALMA sensitivity \textit{remaining constant} as a function of stellar luminosity shows that the exocometary gas release model would naturally produce an increasing occurrence rate for more luminous stars. In other words, the model produces, on average, more high luminosity than low luminosity system above the ALMA CO detection threshold.
This explains the trend of increasing detection rates as a function of stellar luminosity found by \citet{Moor2017}, and visually illustrated by the colors and linestyles of Fig. 3 right (solid green for detected, and dashed red for undetected targets). 


We note that the specific aim of Fig. \ref{fig:predoccurrencerate} is to explain the stellar luminosity trend in the detection rates, and not to compare predicted fluxes to observed fluxes for each of the systems in our sample. The latter was done by
\citet{Kral2017, Kral2018}, who showed that the exocometary gas model can explain all current gas masses observed.
Furthermore, the range of fluxes shown for each belt in Fig. \ref{fig:predoccurrencerate} only accounts for a range of CO(+CO$_2$) mass fractions of 0.8-80\%, consistent with the values derived so far for exocomets and Solar System comets \citep[e.g. Fig. 6 in][]{Matra2017b}. This does not account for other uncertain model parameters which will also influence the flux prediction. For low CO mass belts, the parameter dominating the uncertainty is the unknown molecular excitation for observations of a single transition. For the most massive CO belts, there is also a potential for both extra shielding from CO and \ion{C}{1} prolonging the photodissociation timescale, and for an optical depth greater than 1 for the observed mm transitions, where the latter has been found for some systems through isotopologue observations \citep[e.g.][]{Kospal2013}.

Nonetheless, exocometary gas release is currently the only viable model to explain the presence of CO in low mass systems, since the CO's short photodissociation lifetime requires continuous replenishment. We now showed that the exocometary model can also explain the increasing detection rates as a function of stellar luminosity, as well as the large CO masses observed in some systems \citep{Kral2018}. This gives further support to the idea that CO gas is ubiquitous in planetesimal belts around nearby stars, at least in the cold and bright (and hence mostly young) belts observed so far. 

\section{Conclusions}
\label{sect:concl}
In this work, we presented the first ALMA detection of CO gas in a planetesimal belt around an M star, TWA 7. 
We reach the following conclusions:
\begin{itemize}
    \item TWA 7 hosts $0.8-80\times10^{-6}$ M$_{\oplus}$ of optically thin CO gas, which is consistent with being co-located with the mm dust disk and with being in Keplerian rotation around the star, with the East side moving away from Earth.
    \item The intensity of CO-photodissociating FUV starlight around M dwarfs is dominated by emission from the stellar chromosphere/transition region. Therefore, close to the star where the interstellar radiation field's contribution is negligible, the CO photodissociation rate is sensitively dependent on the overlap of chromospheric lines with the predissociative absorption bands of CO. Young M stars are very active, which means their CO photodissociation rates are, although smaller, not negligible compared to coeval A stars, particularly during flares.
    \item For TWA 7, even assuming a low CO/H$_2$ ratio of 10$^{-6}$, H$_2$ shielding and CO self-shielding could only increase the CO photodissociation timescale from 100 to 1000 years at 60 au. This means that any primordial CO would have been quickly removed, and the observed CO must be produced through exocometary gas release.
    \item Exocometary gas release around an M dwarf through the collisional cascade differs from that around more massive stars. This is because the collisional cascade is affected by stellar wind (as opposed to radiation pressure) -driven removal of the smallest grains, setting the minimum size of the cascade.
    \item We derive a CO(+CO$_2$) mass fraction of TWA 7's exocomets of $\geq70$\%. This is higher than other exocometary gas bearing belts and Solar System comets, and may be explained by 1) uncertainties in the observationally determined minimum size of the cascade, 2) longer photodissociation timescales due to significant shielding of CO by atomic carbon produced as the CO photodissociates, or 3) an intrinsically higher CO(+CO$_2$) content than other belts.
    \item We analyse the stellar luminosity dependence of the occurrence rate of CO gas found by \citet{Moor2017}. We show that this is naturally explained within the framework of our exocometary gas release model, combined with observational detection bias. There is thus no evidence that the presence of CO gas is linked to a specific subset of stars. 
    This supports the idea that CO gas is ubiquitous among the currently surveyed sample of cold, large exocometary belts.
\end{itemize}
Overall, the population of detected belts and the dependence of their occurrence rate on host star luminosity are so far mostly consistent with a picture where exocometary gas is ubiquitous, with all stars releasing it at some level. However, deeper searches in a larger sample of belts, as well as detailed studies of the most CO-rich systems, are necessary to confirm that planetary systems do commonly release exocometary ices throughout their lifetimes.

\acknowledgments
The authors would like to thank G. Kennedy for providing the stellar photospheric model fits.
LM acknowledges support from the Smithsonian Institution as a Submillimeter Array (SMA) Fellow.
A.\,B. and J.\,O. acknowledge financial support from the ICM (Iniciativa Cient\'ifica Milenio) via the N\'ucleo Milenio de Formaci\'on Planetaria grant. J.\,O. acknowledges support from the Universidad de Valpara\'iso, and from Fondecyt (grant 1180395).
This paper makes use of ALMA data ADS/JAO.ALMA\#2015.1.01015.S. ALMA is a partnership of ESO (representing its member states), NSF (USA) and NINS (Japan), together with NRC (Canada), NSC and ASIAA (Taiwan), and KASI (Republic of Korea), in cooperation with the Republic of Chile. The Joint ALMA Observatory is operated by ESO, AUI/NRAO and NAOJ. 



\facility{ALMA}.
\software{CASA \citep[v5.1.0;][]{McMullin2007}, PHOENIX \citep[e.g. ][]{Hauschildt1993}, SciPy \citep{Jones2001}, Matplotlib \citep{Hunter2007}}

\bibliographystyle{apj}
\bibliography{lib}


\end{document}